\documentclass[mlq,fleqn]{w-art}
\usepackage{times}
\usepackage{w-thm}
\usepackage{xspace}
\usepackage{amsmath}
\usepackage{amssymb}
\usepackage{graphicx}
\DeclareFontFamily{U}{mathb}{\hyphenchar\font45}
\DeclareFontShape{U}{mathb}{m}{n}{
      <5> <6> <7> <8> <9> <10> gen * mathb
      <10.95> <12> <14.4> <17.28> <20.74> <24.88> mathb12
      }{}
\DeclareSymbolFont{mathb}{U}{mathb}{m}{n}
\DeclareMathSymbol{\precneq}{3}{mathb}{"AC}
\newcommand{\myiff}{if and only if\xspace}
\newcommand{\reduceq}{\preceq}
\newcommand{\reducneq}{\precneq}

\newcommand{\IR}{\mathbb{R}}
\newcommand{\IRc}{\mathbb{R}_{\text{c}}}

\newcommand{\IQ}{\mathbb{Q}}

\newcommand{\IN}{\mathbb{N}}
\newcommand{\calA}{\mathcal{A}}
\newcommand{\calS}{\mathcal{S}}

\newcommand{\cf}[1]{\mathbf{1}_{#1}}
\newcommand{\dist}[1]{\operatorname{dist}_{#1}}

\newcommand{\Semiequiv}{\makebox[0pt][l]{\raisebox{-0.6ex}{$\nLeftarrow$}}\hspace{0.5ex}\raisebox{0.5ex}{$\Rightarrow$}}%
\newcommand{\calO}{\mathcal{O}}

\newcommand{\person}[1]{\textsc{#1}}
\newcommand{\aname}[1]{\textsf{#1}}
\newcommand{\mycite}[2]{{\rm\cite[\textsc{#1}]{#2}}}
\newcommand{\myto}{\!\to\!}

\newcommand{\ball}{B}
\newcommand{\closure}[1]{\overline{#1}}

\newcommand{\cball}{\closure{\ball}}
\newcommand{\oball}{\ball^\circ}
\newcommand{\myrho}{\rho}
\newcommand{\myl}{{\scriptscriptstyle<}}
\newcommand{\myg}{{\scriptscriptstyle>}}
\newcommand{\myrhol}{\myrho_{\raisebox{0.2ex}{$\myl$}}}
\newcommand{\myrhog}{\myrho_{\raisebox{0.2ex}{$\myg$}}}

\newcommand{\psiL}[1]{\psi^{\hspace*{-0.7pt}#1}_{\!\raisebox{0.2ex}{$\myl$}}}
\newcommand{\psiG}[1]{\psi^{\hspace*{-0.7pt}#1}_{\!\raisebox{0.2ex}{$\myg$}}}

\newcommand{\thetaL}[1]{\theta^{#1}_{\raisebox{0.5ex}{$\scriptscriptstyle<$}}}

\newcommand{\psil}{\psi_{\!\raisebox{0.2ex}{$\myl$}}}
\newcommand{\psig}{\psi_{\!\raisebox{0.2ex}{$\myg$}}}

\newcommand{\continuum}{\mathfrak{c}}

\newcommand{\Card}[1]{\operatorname{Card}({#1})}
\newcommand{\COMMENTED}[1]{}

\begin{document}
\Volume{49}
\Issue{49}
\Month{01}
\Year{2003}
\pagespan{1}{}
\title[Singular Coverings and Non-Uniform Notions of Closed Set Computability]{%
Singular Coverings and \\ Non-Uniform Notions of Closed Set Computability}
\author[Le Roux]{St\'{e}phane Le Roux\footnote{%
e-mail: {\sf stephane.le.roux@ens-lyon.fr};
supported by the
\emph{Minist\`{e}re des Affaires Etrang\`{e}res} with scholarship {\sf CDFJ},
with {\sf Explo`ra doc} from the \emph{R\'{e}gion Rh\^{o}ne-Alpes},
and by the \emph{Minist\`{e}re de l'Enseignement Sup\'{e}rieur et de
la Recherche.}}%
\inst{1,2}}%
\author[Ziegler]{Martin Ziegler\footnote{%
e-mail: {\sf ziegler@uni-paderborn.de};
supported by the \emph{Japanese Society for the
Promotion of Science} (\textsf{JSPS}) grant {\sf PE\,05501} and
by the \emph{German Research Foundation} (DFG) project {\sf Zi\,1009/1-1}.}%
\inst{1,3}}
\address[\inst{1}]{Japan Advanced Institute of Science and Technology}
\address[\inst{2}]{Ecole normale sup\'{e}rieure de Lyon}
\address[\inst{3}]{University of Paderborn}
\begin{abstract} 
The empty set of course contains no computable point.
On the other hand, surprising results due to 
Zaslavski\u{\i}, Tse\u{\i}tin, Kreisel, and Lacombe
have asserted the existence of \emph{non-}empty co-r.e. closed 
sets devoid of computable points: sets 
which are even `large' in the sense of positive
Lebesgue measure. 

This leads us to investigate
for various classes of 
computable real subsets 
whether they necessarily 
contain a (not necessarily effectively findable) 
computable point. 
\end{abstract}
\maketitle
\newtheorem{observation}[theorem]{Observation}
\newtheorem{fact}[theorem]{Fact}
\newtheorem{scholium}[theorem]{Scholium}
\newtheorem{scholiumf}[theorem]{Scholium\footnotemark}
\newtheorem{propositionf}[theorem]{Proposition\footnotemark}
\newtheorem{question}[theorem]{Question}
\section{Introduction} \label{s:Intro}
A discrete set $A$, for example a subset of $\{0,1\}^*$ or $\IN$,
is naturally called r.e. (i.e. semi-decidable)
if a Turing machine can enumerate the members of
(equivalently: terminate exactly for inputs from) $A$.
The corresponding notions for open subsets of reals
\cite{LacombeI,LacombeII,Weihrauch} amount to the following

\begin{definition}\it \label{d:Open} 
Fix a dimension $d\in\IN$.
An open subset $U\subseteq\IR^d$ is called
\emph{r.e.} ~\myiff~ a Turing machine can enumerate
rational centers $\vec q_n\in\IQ^d$ and radii $r_n\in\IQ$
of open Euclidean balls 
$\oball(\vec q,r)=\big\{\vec x\in\IR^d:\|\vec x-\vec q\|<r\big\}$
exhausting $U$.

A real vector $\vec x\in\IR^d$ is (Cauchy-- or $\myrho^d$--)\emph{computable}
~\myiff~ a Turing machine can generate a sequence $\vec q_n\in\IQ^d$
of rational approximations converging to $\vec x$ \emph{fast}
in the sense that $\|\vec x-\vec q_n\|\leq2^{-n}$.
\end{definition}
\noindent
Notice that an open real subset is r.e. ~\myiff~
membership ``$\vec x\in U$'' is semi-decidable
with respect to $\vec x$ given by fast convergent rational
approximations; see for instance \cite[\textsc{Lemma~4.1}c]{MLQ2}.

\subsection{Singular Coverings} \label{s:Singular}
A surprising result due to \person{E.~Specker} 
implies that the (countable) set $\IRc$ of computable reals
is contained in an r.e. open \emph{proper} subset
$U$ of $\IR$: In his work \cite{Specker} he
constructs a computable function $f:[0,1]\to[0,\tfrac{1}{36}]$
attaining its maximum $\tfrac{1}{36}$ in no computable point;
hence $U:=(-\infty,0)\cup f^{-1}[(-1,\tfrac{1}{36})]\cup(1,\infty)$ 
has the claimed properties,
see for example \mycite{Theorem~6.2.4.1}{Weihrauch}.
This was strengthened in \cite{Zaslavski,Kreisel}
to the following

\begin{fact} \label{f:Singular}
For any $\epsilon>0$ there exists an r.e. open
set $U_{\epsilon}\subseteq\IR$ of Lebesgue measure $\lambda(U_\epsilon)<\epsilon$
containing all computable real numbers.
\end{fact}
\begin{proof}
See \mycite{Section~8.1}{Kushner} or
\mycite{Section~IV.6}{Beeson} or
\mycite{Theorem~4.2.8}{Weihrauch}.
\end{proof}
The significance of this improvement thus lies in the
constructed $U_\epsilon$ intuitively being very `small':
it misses many non-computable points. On the other hand
it is folklore 
that a certain smallness is also necessary: Every r.e. open
$U\subsetneq\IR$ covering $\IRc$ \emph{must} miss uncountably
many non-computable points.
Put differently, an at most countable non-empty closed real subset
must, if its complement is r.e., contain a computable point;
see Observation~\ref{o:Main} below.

This leads the present work to study further natural
effective classes of closed Euclidean sets with respect
to the question whether they contain a computable point.
But let us start with reminding of the notion of

\section{Computability of Closed Subsets} \label{s:Closed}
Decidability of a discrete set $A\subseteq\IN$
amounts to computability of its characteristic function
\[ \cf{A}(x) \;=\; 1   \quad\text{if}\quad x\in A, \qquad
   \cf{A}(x) \;=\; 0   \quad\text{if}\quad x\not\in A \enspace . \]
Literal translation to the real number setting fails
of course due to the continuity requirement; instead,
the characteristic function is replaced by the continuous distance function 
\[ \dist{A}(x) \;=\; \inf\big\{ \|x-a\| : a\in A\big\} \]
which gives rise to the following natural notions \cite{Closed},
\mycite{Corollary~5.1.8}{Weihrauch}:

\begin{definition}\it \label{d:Closed}
Fix a dimension $d\in\IN$.
A closed subset $A\subseteq\IR^d$ is called
\begin{itemize}
\item[\textbullet]
  \emph{r.e.} \ \myiff \ $\dist{A}:\IR^d\to\IR$ 
  is upper computable;
\item[\textbullet]
  \emph{co-r.e.} \ \myiff \ $\dist{A}:\IR^d\to\IR$ 
  is lower computable;
\item[\textbullet]
  \emph{recursive} \ \myiff \ $\dist{A}:\IR^d\to\IR$ 
  is computable.
\end{itemize}
\end{definition}
\noindent
Lower computing $f:\IR^d\to\IR$ 
amounts to the output, given a sequence $(\vec q_n)\in\IQ^d$
with $\|\vec x-\vec q_n\|\leq2^{-n}$, 
of a sequence $(p_m)\in\IQ$ with $f(\vec x)=\sup_m p_m$.
This intuitively means approximating $f$ from below and is
also known as $(\myrho^d,\myrhol)$--computability
with respect to standard real representations $\myrho$ and $\myrhol$;
confer \mycite{Section~4.1}{Weihrauch} or \cite{SemiTCS}.
A closed set is co-r.e. ~\myiff~ its complement (an open set)
is r.e. in the sense of Definition~\ref{d:Open}
\mycite{Section~5.1}{Weihrauch}.
Several other reasonable notions of closed set computability have turned out 
as equivalent to one of the above;
see \cite{Closed} or \mycite{Section~5.1}{Weihrauch}:
recursivity for instance is equivalent to \emph{Turing location} \cite{Ge}
as well as to being simultaneously r.e. and co-r.e.
This all has long confirmed Definition~\ref{d:Closed} as natural indeed.

\subsection{Non-Empty Co-R.E. Closed Sets without Computable Points} 
\label{s:ComputablePoints}
Like in the discrete case, r.e. and co-r.e. are logically independent
also for closed real sets:

\begin{example} \label{x:Continuum}
For $x:=\sum_{n\in H}2^{-n}$ (where $H\subseteq\IN$ denotes the Halting Problem),
the compact interval $I_{\myl}:=[0,x]\subseteq\IR$ is r.e. but not co-r.e.;
and $I_{\myg}:=[x,1]$ is co-r.e. but not r.e.
\qed\end{example}
\noindent
Notice that both intervals have continuum cardinality
and include lots of computable points.
As a matter of fact, it is a well-known
\begin{fact} \label{f:positive}
Let $A\subseteq\IR^d$ be r.e. closed and non-empty.
Then $A$ contains a computable point
{\rm\cite[\textsc{Exercise~5.1.13}b]{Weihrauch}}.
\end{fact}
More precisely, closed $\emptyset\not=A\subseteq\IR^d$ 
is r.e. ~\myiff~ $A=\closure{\{\vec x_1,\ldots,\vec x_n,\ldots\}}$
for some computable sequence $(\vec x_n)_{_n}$ of real vectors
\mycite{Lemma~5.1.10}{Weihrauch}.

A witness of (one direction of) logical independence 
stronger than $I_{\myg}$ is thus a non-empty co-r.e. closed
set $A$ devoid of computable points: $A\subseteq[0,1]\setminus\IRc$.
For example every singular covering $U_\epsilon$ with $\epsilon<1$ from Section~\ref{s:Singular} 
due to \cite{Zaslavski,Kreisel} gives rise to an instance $A_\epsilon:=[0,1]\setminus U_\epsilon$
even of positive Lebesgue measure $\lambda(A)>1-\epsilon$, and thus of continuum cardinality.
Conversely, it holds

\begin{observation} \label{o:Main}
Every non-empty co-r.e. closed set of cardinality
strictly \emph{less} than that of the continuum \emph{does} 
contain computable points.
\end{observation}
Notice that this claim also covers 
putative cardinalities between
$\aleph_0$ and $2^{\aleph_0}=\continuum$
i.e. does not rely on the \textsf{Continuum Hypothesis}.

In a finite set, every point is isolated;
in this case the claim thus follows from
the well-known

\begin{fact} \label{f:Isolated}
\begin{enumerate}
\item[a)]
Let $A\subseteq\IR^d$ be co-r.e. closed
and suppose there exist $\vec a,\vec b\in\IQ^d$
such that $A\cap[\vec a,\vec b]=\{\vec x\}$
(where $[\vec a,\vec b]:=\prod_{i=1}^d [a_i,b_i]$).
Then, $\vec x$ is computable.
\item[b)]
A \textsf{perfect} subset $A\subseteq X$ (of $X=\IR^d$ or of $X=\{0,1\}^\omega$),
i.e. one which coincides with the collection $A'$ of its limit points,
\[ A' \quad:=\quad \big\{ \vec x\in X \;\big|\; 
 \forall n \exists \vec a\in A : \; 0<|\vec a-\vec x|<1/n \big\} 
\enspace , \]
is either empty or of continuum cardinality.
\end{enumerate}
\end{fact}
See for instance \mycite{Proposition~3.6}{Closed} and 
\mycite{Corollary~6.3}{Kechris}.

\begin{proof}[Proof (Observation~\ref{o:Main}).]
Suppose that $A$ has cardinality strictly less than that of the continuum.
Then $A\not=A'$ by Fact~\ref{f:Isolated}b).
On the other hand, $A$ contains $A'$ because it is closed.
Hence the difference $A\setminus A'\not=\emptyset$ holds
and consists of isolated points which 
are computable by Fact~\ref{f:Isolated}a).
\end{proof}
\noindent
So every non-empty co-r.e. closed real set 
$A\subseteq[0,1]$ devoid of computable points
must necessarily be of continuum cardinality. On the other hand, 
Fact~\ref{f:Singular} yields 
such sets with positive Lebesgue measure
$\lambda(A)>0$. In view of (and in-between) the strict\footnote{Consider
for instance the irrational numbers $\IR\setminus\IQ$ and 
\person{Cantor}'s uncountable \textsf{Middle Third} set, respectively.}
chain of implications
\[
\text{nonempty interior} \quad\Semiequiv\quad
\text{positive measure} \quad\Semiequiv\quad
\text{continuum cardinality} \]
we make the following\footnote{We are grateful to a careful anonymous referee
for indicating this simple solution to a question raised in an
earlier version of this work.}
\begin{remark} \label{r:Cenzer} \it
There exists a non-empty co-r.e. closed real subset of measure zero
without computable points.
\end{remark}
\noindent
This is different from \mycite{Section~8.1}{Kushner} which considers
\begin{itemize}
\item[--] coverings of $(0,1)$ having measure \emph{strictly} less than 1
\item[--] by disjoint enumerable `segments', that is \emph{closed} intervals $[a_n,b_n]$,
\item[--] or by enumerable open intervals $(a_n,b_n)$ as in Definition~\ref{d:Open},
 however in terms of the \emph{accumulated} length $\sum_n (b_n-a_n)$,
 that is counting interval overlaps doubly \mycite{Theorem~8.5}{Kushner}.
\end{itemize}
\begin{proof}[Proof (Remark~\ref{r:Cenzer}).]
Take a subset $A$ of Cantor space with these properties 
and consider its image $\tilde A$ under the canonical embedding 
\[ \{0,1\}^\omega \;\ni\; (b_n) \;\mapsto\; \sum_n b_n 2^{-n} \;\in\; [0,1] \enspace . \]
Notice that this mapping, restricted to $A$, is indeed injective
because only dyadic rationals have a non-unique binary expansion;
and in fact two of them, both of which are decidable. Therefore
\begin{itemize}
\item $\tilde A$ has continuum cardinality but, being contained
  in Cantor's \textsf{Middle Third} set, has measure zero.
\item The enumeration of open balls in $\{0,1\}^\omega$ exhausting
  $A$'s complement translates to one exhausting $[0,1]\setminus\tilde A$.
\item Suppose $x\in\tilde A$ were computable. Then $x$ has 
  decidable binary expansion \mycite{Theorem~4.1.13.2}{Weihrauch},
  contradicting that all elements of $\tilde A$ arise from
  uncomputable binary sequences $(b_n)\in A$.
\end{itemize}\end{proof}

\subsection{Computability on Classes of Closed Sets of Fixed Cardinality} \label{s:Nonuniform}
Observation~\ref{o:Main} and Fact~\ref{f:Isolated}a) 
are non-uniform claims: they assert a computable point
in $A$ to \emph{exist} but not that it can be `found'
effectively.
Nevertheless, a uniform version of Fact~\ref{f:Isolated}a) 
does hold under the additional hypothesis
that $\vec a$ and $\vec b$ are known;
compare \mycite{Exercise~5.2.3}{Weihrauch}
reported as Lemma~\ref{l:Finite}a) below.
The present section investigates whether
and to what extend this result can be generalized
towards Observation~\ref{o:Main} and, to this end,
considers the following representations for
(classes of) closed real sets of fixed cardinality:

\begin{definition}\it \label{d:Representations}
For $d\in\IN$ and closed $A\subseteq\IR^d$, 
\begin{itemize}
\itemsep4pt
\item[\textbullet]
  $\psiL{d}$ encodes $A$ as a $[\myrho^d\myto\myrhog]$--name of $\dist{A}$;
\item[\textbullet]
  $\psiG{d}$ encodes $A$ as a $[\myrho^d\myto\myrhol]$--name of $\dist{A}$
\end{itemize}
in the sense of {\rm\cite{SemiTCS}}.

Write
$\calA^d_N:=\{A\subseteq[0,1]^d\text{ closed}:\Card{A}=N\}$
for the hyperspace of compact sets having cardinality exactly $N$,
where $N\leq\continuum$ denotes a cardinal number.
Equip $\calA^d_N$ with restrictions
$\psiL{d}|^{\calA^d_N}$ and $\psiG{d}|^{\calA^d_N}$
of the above representations.

If $N\leq\aleph_0$, we furthermore can encode $A\subseteq[0,1]^d$
of cardinality $N$ (closed or not) by the join of the $\myrho^d$--names
of the $N$ elements constituting $A$, listed in arbitrary order\footnote{%
see also Lemma~\ref{l:Countable}a)}.
This representation shall be denoted as $(\myrho^d)^{\sim N}$.
\end{definition}
\noindent
Let us first handle finite cardinalities:
\begin{lemma} \label{l:Finite}
Fix $d\in\IN$. 
\begin{enumerate}
\itemsep4pt
\item[a)] 
  $\displaystyle\psiL{d}\big|^{\calA^d_1}\;\equiv\;(\myrho^d)^{\sim 1}\;\equiv\;\psiG{d}\big|^{\calA^d_1}$
\item[b)] 
  For $2\leq N\in\IN$, it holds
  $\displaystyle\psiL{d}\big|^{\calA^d_N}\;\equiv\;(\myrho^d)^{\sim N}\;\reducneq\;\psiG{d}\big|^{\calA^d_N}$
\item[c)]
  For $N\in\IN$,
  $A\in\calA^d_N$ is ~$\psiL{d}$--computable ~\myiff~ it is $\psiG{d}$--computable.
\end{enumerate}
\end{lemma}
\noindent
In particular, \mycite{Example~5.1.12.1}{Weihrauch}
generalizes to arbitrary finite sets:

\begin{corollary} \label{c:Finite}
A finite subset $A$ of $\IR^d$ is 
r.e. ~\myiff~ $A$ is co-r.e. ~\myiff~ every point in $A$ is computable.
\end{corollary}

\begin{proof}[Proof (Lemma~\ref{l:Finite}).]
\begin{enumerate}
\item[a)]
Confer \mycite{Exercise~5.2.3}{Weihrauch}.
\item[b)]
The reductions ``$(\myrho^d)^{\sim N}\reduceq\psiL{d}$'' 
and ``$(\myrho^d)^{\sim N}\reduceq\psiG{d}$'' follow
from induction on $N$ via Claim~a) and
\mycite{Theorem~5.1.13.1}{Weihrauch}.
``$\psig|^{\calA_2}\reducneq(\myrho)^{\sim 2}$'' can be seen easily
based on a straight-forward discontinuity argument.
For ``$\psiL{d}|^{\calA^d_N}\reduceq(\myrho^d)^{\sim N}$'',
recall that a $\psiL{d}$--name for $A$ is (equivalent to) the
$\myrho^d$--names of countably infinitely many points $\vec x_m\in A$
dense in $A$ \mycite{Lemma~5.1.10}{Weihrauch}.
Since $A=\{\vec a_1,\ldots,\vec a_N\}$ is finite,
there exist $m_1,\ldots,m_N\in\IN$ such that 
$\vec x_{m_n}=\vec a_n$ for $n=1,\ldots,N$;
equivalently: $\vec x_{m_i}\not=\vec x_{m_j}$ for $i\not=j$.
The latter condition also yields a way to effectively
find such indices $m_1,\ldots,m_N$, regarding that
\emph{in}equality is semi-decidable.
Once found, the 
$\myrho^d$--names of $\vec x_{m_1},\ldots,\vec x_{m_N}$
constitute a $(\myrho^d)^{\sim N}$--name of $A$.
\item[c)]
That $\psiL{d}$--computability implies $\psiG{d}$--computability follows from a).
For the converse, the case $N=1$ is covered in Claim~a).
In case $N>1$ exploit that points of $A$ lie isolated;
that is, there exist closed rational cubes 
$Q_n:=[\vec a_n,\vec b_n]$,
$n=1,\ldots,N$ containing exactly one element of $A$ each.
By storing their finitely many coordinates, a Type-2 machine is
able to $\psiG{d}$--compute the $N$ closed one-element sets $A\cap Q_n$.
We have thus effectively reduced to the case $N=1$.
\end{enumerate}\end{proof}
The case of countably infinite closed sets:

\begin{lemma} \label{l:Countable}
\begin{enumerate}
\item[a)]
  In the definition of $(\myrho^d)^{\sim\aleph_0}$, it does not
  matter whether each element $\vec x$ of $A$ is required to
  occur exactly once or at least once.
\item[b)] 
  It holds \ 
  $\displaystyle(\myrho^d)^{\sim\aleph_0}\big|^{\calA^d_{\aleph_0}}
   \;\reducneq\;\psiL{d}\big|^{\calA^d_{\aleph_0}}$. 
\item[c)]
  There exists a countably infinite r.e. closed set $A\subseteq[0,1]$
  which is neither $\myrho^{\sim\aleph_0}$--computable
  nor co-r.e.
\item[d)]
  There is a countably infinite co-r.e. 
  but not r.e.
  closed set $B\subseteq[0,1]$.
\end{enumerate}
\end{lemma}
\begin{proof}\begin{enumerate}
\item[a)]
  \mycite{Exercise~4.2.3}{Weihrauch} holds uniformly
  and extends from $\IR$ to $\IR^d$:
  Let vectors $\vec x_m\in\IR^d$ be given by $\myrho^d$--names, $m\in\IN$.
  Based on semi-decidable \emph{in}equality ``$\vec x_m\not=\vec x_n$'',
  we can employ dove-tailing to identify infinitely many distinct ones
  like in the proof of Lemma~\ref{l:Finite}a).
  However infinitely many may be not enough:
  some care is required to find \emph{all} of them. To this end, take the
  given $\vec q_{m,n}\in\IQ^d$ with $\|\vec q_{m,n}-\vec x_m\|_2<2^{-n}$
  and suppose $M_n\subseteq\IN$ is a finite set of 
  indices of vectors already identified as distinct, that is,
  with $\vec x_m\not=\vec x_{m'}$ for $m,m'\in M_n$, $m\not=m'$.
  Then, in phase $n+1$, consider the smallest (!) index $m_{n+1}$
  newly recognized as different from all $\vec x_m$, $m\in M_n$:
\begin{gather*}
  M_n'\;:=\;\big\{ m'\in\IN\setminus M_n \;\big|\;
  \forall m\in M_n: \;  \oball\big(\vec q_{m,n},2^{-n}\big) \:\cap\:
  \oball\big(\vec q_{m',n},2^{-n}\big) \:=\emptyset\big\}, \\[0.5ex]
   m_{n+1} \;:=\; \min M_n',
  \qquad   
M_{n+1}:=M_n\cup\{m_{n+1}\}
\end{gather*}
  if $M_n'\not=\emptyset$, otherwise $M_{n+1}:=M_n$.
  Notice that an element of $M_n'$ re-appears
  in $M_{n+1}'$ unless it was the minimal one:
  if balls are disjoint, they remain so when reducing the radius.
  Let's argue further to assert correctness of this algorithm:
  By prerequisite there are infinitely many distinct vectors
  among the $(\vec x_m)$, hence $M_n'\not=\emptyset$ 
  infinitely often, yielding a
  sequence $(\vec x_{m_n})_n$ (not a \emph{sub}sequence since
  that would require $(m_n)$ to be increasing) of 
  distinct elements; in fact of \emph{all} of them:
  If $m'\in\IN$ is such that $\vec x_{m'}\not=\vec x_m$
  for all $m<m'$, then there exists some 
  $n\in\IN$ for which $m'\in M_{n}'$.
  By virtue of the above observation,  $m'$
  eventually becomes the minimal element
  of some later $M_{n'}'$ and thus does occur in the output
  as index $m_{n'}$.
\item[b)]
  The positive part of the claim
  follows from \mycite{Lemma~5.1.10}{Weihrauch},
  asserting that a sequence of real vectors dense in closed $A$
  yields a $\psil$--name of $A$;
  whereas (negative) unreducibility is a consequence of Claim~c).
\item[c)]
  Let $(x_n)_{_n}$ denote a Specker Sequence,
  that is, a computable sequence converging (non-effectively)
  from below to the uncomputable real $x_\infty=\sum_{n\in H}2^{-n}$.
  Then $(x_n)_{_n}$ is dense in closed 
  $A:=\{\vec x_n:n\in\IN\}\cup\{x_\infty\}\subseteq[0,1]$,
  hence the latter $\psil$--computable.
  But $x_\infty=\max A$ is $\myrhog$--uncomputable,
  therefore $A$ cannot be $(\myrho)^{\sim\aleph_0}$--computable;
  nor $\psig$--computable \mycite{Lemma~5.2.6.2}{Weihrauch}.
\item[d)]
  For the Halting Problem $H\subseteq\IN$
  consider the closed set $B:=\{0\}\cup\{2^{-n}:n\not\in H\}\subseteq[0,1]$.
  The open rational intervals $(2^{-n-1},2^{-n})$ for all $n\in\IN$
  and, for $n\in H$ by semi-decidability, $(2^{-n-1},3\cdot2^{-n-1})$
  together 
  exhaust exactly $(0,1)\setminus A$; this enumeration thus establishes
  $\psig$--computability of $B$. $\psil$--computability 
  fails due to \mycite{Exercise~5.1.5}{Weihrauch}.
\end{enumerate}\end{proof}

\section{Closed Sets and Naively Computable Points} \label{s:Naive}
A notion of real computability weaker than that of 
Definition~\ref{d:Open} is given in the following

\begin{definition}\it \label{d:Naive}
A real vector $\vec x\in\IR^d$ is \emph{naively computable}
(also called \emph{recursively approximable})
~if~ a Turing machine can generate a sequence $\vec q_n\in\IQ^d$
with $\vec x=\lim_n\vec q_n$ (i.e. converging but not necessarily fast).
\end{definition}
\noindent
A real point is naively computable
~\myiff~ it is Cauchy--computable \emph{relative} to the Halting oracle $H=\emptyset'$,
see \mycite{Theorem~9}{Ho} or \cite{ArithHierarchy}.

Section~\ref{s:ComputablePoints} asked whether \emph{certain}
non-empty co-r.e. closed sets contain a Cauchy--computable element.
Regarding naively computable elements, it holds

\begin{propositionf}\footnotetext{A simple reduction to
the counterpart of this claim 
for Baire space \cite[\textsc{Theorem~2.6}(c)]{Cenzer} 
does not seem feasible because, according to
\mycite{Theorem~4.1.15.1}{Weihrauch}, there exists no 
\emph{total} (compact or not) representation 
equivalent to $\myrho$.} \label{p:Naive}
\emph{Every} non-empty co-r.e. closed set $A\subseteq\IR^d$
contains a naively computable point $\vec x\in A$.
\end{propositionf}
W.l.o.g. $A$ may be presumed compact by proceeding to
$A\cap[\vec u,\vec v]$ for appropriate 
$\vec u,\vec v\in\IQ^d$ \mycite{Theorem~5.1.13.2}{Weihrauch}.
In 1D one can then explicitly choose $x=\max A$ according 
to \mycite{Lemma~5.2.6.2}{Weihrauch}. For higher dimensions
we take a more implicit approach and apply
Lemma~\ref{l:neg2pos}a) to
the following relativization of Fact~\ref{f:positive}:
\begin{scholiumf}\footnotetext{A \aname{scholium} is ``\emph{a 
note amplifying a proof or course of reasoning,
as in mathematics}'' \cite{Dictionary}}
Let non-empty $A\subseteq\IR^d$ be r.e. closed 
\emph{relative} to $\calO$ for some oracle $\calO$.
Then $A$ contains a point computable \emph{relative} to $\calO$.
\end{scholiumf}
\begin{lemma} \label{l:neg2pos}
Fix closed $A\subseteq\IR^d$.
\begin{enumerate}
\item[a)]
If $A$ is co-r.e.,
then it is also r.e. relative to $\emptyset'$.
\item[b)]
If $A$ is r.e.,
then it is also co-r.e. relative to $\emptyset'$.
\end{enumerate}
\end{lemma}
These claims may follow
from \cite{Borel,Gherardi}. However for purposes
of self-containment we choose to give a direct
\begin{proof}
Recall \mycite{Definition~5.1.1}{Weihrauch} that
a $\psiG{d}$--name of $A$ is an enumeration of all
closed rational balls $\cball$ disjoint from $A$;
whereas a $\psiL{d}$--name enumerates all
open rational balls $\oball$ intersecting $A$.
Observe that
\begin{equation} \label{e:pos2neg}
\begin{aligned}
\oball\cap A\not=\emptyset &\quad\Leftrightarrow\quad
  \exists n\in\IN: \cball_{\scriptscriptstyle -1/n}\cap A\not=\emptyset
\\
\cball\cap A=\emptyset &\quad\Leftrightarrow\quad
  \exists n\in\IN: \oball_{\scriptscriptstyle +1/n}\cap A=\emptyset
\end{aligned}
\end{equation}
where $\ball_{\pm\epsilon}$ means enlarging/shrinking
$\ball$ by $\epsilon$ such that
$\oball=\bigcup_n\cball_{+1/n}$ and
$\cball=\bigcap_n\oball_{-1/n}$. Formally in 1D e.g.
$(u,v)_{-\epsilon}:=(u+\epsilon,v-\epsilon)$ in case
$v-u>2\epsilon$, $(u,v)_{-\epsilon}:=\{\}$ otherwise.
Under the respective hypothesis of a) and b),
the corresponding right hand side of Equation~(\ref{e:pos2neg})
is obviously decidable relative to $\emptyset'$.
\end{proof}
\noindent
A simpler argument might try to exploit 
\mycite{Theorem~9}{Ho}
that every $\myrhol$--computable single real $y$ is,
relative to $\emptyset'$, $\myrhog$--computable;
and conclude by uniformity that (Definition~\ref{d:Closed})
every $(\myrho\myto\myrhol)$--computable
function $f:x\mapsto f(x)=y$ is,
relative to $\emptyset'$, $(\myrho\myto\myrhog)$--computable.
This conclusion
however is \underline{wrong} in general because even
a relatively $(\myrho\myto\myrhog)$--computable $f$
must be upper semi-continuous
whereas a $(\myrho\myto\myrhol)$--computable
one may be merely lower semi-continuous.

\subsection{(In-)Effective Compactness}
By virtue of the \textsf{Heine--Borel} and \textsf{Bolzano--Weierstrass} Theorems, 
the following properties of a real subset $A$ are equivalent:
\begin{enumerate}
\item[i)] $A$ is closed and bounded;
\item[ii)] every open rational cover $\bigcup_{n\in\IN}\oball(\vec q_n,r_n)$
 of $A$ contains a finite sub-cover;
\item[iii)] any sequence $(\vec x_n)$ in $A$ 
 admits a subsequence $(\vec x_{n_k})$
 converging within $A$.%
\end{enumerate}
Equivalence~ ``i)$\Leftrightarrow$ii)'' (Heine--Borel) carries over
to the effective setting \mycite{Lemma~5.2.5}{Weihrauch}
\mycite{Theorem~4.6}{Closed}. Regarding 
\textsf{sequential compactness} iii), a Specker Sequence
(compare the proof of Lemma~\ref{l:Countable}c) yields the
counter-example of a recursive rational sequence in $A:=[0,1]$
having no recursive \emph{fast} converging subsequence,
that is, no computable accumulation point.
This leaves the question whether every bounded recursive
sequence admits an at least \emph{naively} computable 
accumulation point. Simply taking the \emph{largest} one 
(compare the proof of Proposition~\ref{p:Naive} in case $d=1$) 
does not work in view of \mycite{Theorem~6.1}{ArithHierarchy}.
Also effectivizing the Bolzano--Weierstra\ss{} 
selection argument yields only an 
accumulation point computable relative to $\emptyset'\pmb{'}$:

\begin{observation} \label{o:Sigma3}
Let $(x_n)\subseteq[0,1]$ be a bounded sequence.
For each $m\in\IN$ choose $k=k(m)\in\IN$ such that
there are infinitely many $n$ with $x_n\in\oball(x_k,2^{-m})$.
Boundedness and pigeonhole principle,
inductively for $m=1,2,\ldots$, assert the existence 
of smaller and smaller (length $2^{-m}$) sub-intervals 
each containing infinitely many members of that sequence:
\begin{equation} \label{e:Sigma3}
\exists a,b\in\IQ \;\;\forall N\;\; \exists n\geq N: 
\quad x_n\in(a,b)\;\wedge\;|b-a|\leq2^{-m}
\enspace . \end{equation}
This is a $\Sigma_3$--formula; and thus
semi-decidable \emph{relative} to $\emptyset''$,
see for instance
\mycite{Post's Theorem~\textsection IV.2.2}{Soare}.
\end{observation}
In fact $\emptyset''$ is the best possible as 
we establish, based on Section~\ref{s:ratHP},
\begin{theorem} \label{t:Lagnese}
There exists a recursive rational sequence $(x_n)\subseteq[0,1]$
containing no \emph{naively} computable accumulation point.
\end{theorem}
\noindent
This answers a recent question in \textsf{Usenet} \cite{Lagnese}.
The sequence constructed is rather complicated---and
must be so in view of the following counter-part to
Fact~\ref{f:Isolated}a) and Observation~\ref{o:Main}:
\begin{lemma}
Let $(x_n)\subseteq[0,1]^d$ be a computable real sequence
and let $A$ denote the set of its accumulation points.
\begin{enumerate}
\item[a)]
  Every isolated point $x$ of $A$ is naively computable.
\item[b)]
  If $\Card{A}<\continuum$, then $A$ contains
  a naively computable point.
\end{enumerate}
\end{lemma}
\begin{proof}
$A$ is closed non-empty
and thus, if in addition free of isolated points,
perfect; so b) follows from a).
Let $\{x\}=A\cap[u,v]=A\cap(r,s)$ with rational $u<r<s<v$.
A subsequence $(x_{n_m})$ contained in $(r,s)$ 
will then necessarily converge to $x$. 
Naive computability of $x$ thus follows from
selecting such a subsequence effectively:
Iteratively for $m=1,2,\ldots$ use dove-tailing to
search for (and, as we know it exists, also find)
some integer $n_m>n_{m-1}$ with ``$x_{n_m}\in(r,s)$''.
The latter property is indeed semi-decidable, for instance by virtue of
\cite[\textsc{Lemma~4.1}c]{MLQ2}.
\end{proof}
\noindent
We have just been pointed out \cite{Xizhong}
that Theorem~\ref{t:Lagnese}
can easily be proven by a standard diagonalization on
an enumeration of all recursive rational sequences.
However we prefer an alternative approach because
the uniform Proposition~\ref{p:ratHP} below may be
of interest of its own. Indeed, Theorem~\ref{t:Lagnese} 
follows from applying to Proposition~\ref{p:ratHP} 
a relativization of Fact~\ref{f:Singular} which
is an easy consequence of for example the 
proof of \mycite{Theorem~4.2.8}{Weihrauch},
namely

\begin{scholium}
For any oracle $\calO$,
there exists a non-empty closed set $A\subseteq[0,1]$
co-r.e. \emph{relative} to $\calO$,
containing no point Cauchy--computable \emph{relative} to $\calO$.
\end{scholium}

\subsection{Co-R.E. Closed Sets Relative to \boldmath${\emptyset'}$} \label{s:ratHP}
\mycite{Theorem~9}{Ho} has given a nice characterization
of real numbers Cauchy--computable \emph{relative} to the Halting oracle.
We do similarly for co-r.e. closed real sets:

\begin{proposition} \label{p:ratHP}
A closed subset $A\subseteq\IR^d$
is $\psiG{d}$--computable \emph{relative} to $\emptyset'$
~\myiff~ it is the set of accumulation points
of a recursive rational sequence or,
equivalently, of an enumerable infinite subset of rationals.
\end{proposition}
\noindent
This follows (uniformly and for simplicity in case $d=1$) from Claims~a-e) of%

\begin{lemma} \label{l:ratHP}
\begin{enumerate}
\item[a)]
Let closed $A\subseteq\IR$ be co-r.e. relative to $\emptyset'$.
Then there is a recursive double sequence of open rational 
intervals $\oball_{m,n}=(u_{m,n},v_{m,n})$
and a (not necessarily recursive) function $M:\IN\to\IN$ such that
\begin{enumerate}
\item[i)] $\forall N\in\IN\; \forall m\geq M(N)\; \forall n\leq N:  
  \quad \oball_{m,n}=\oball_{M(N),n}=\ldots=:\oball_{\infty,n}$ ~ 
  ($\oball_{m,1},\ldots,\oball_{m,N}$ 
  each stabilizes beyond $m\geq M(N)$)
\item[ii)] $A=\IR\setminus\bigcup_n \oball_{\infty,n}$.
\end{enumerate}
\item[b)]
From a double sequence $\oball_{m,n}$ of open rational intervals as in a\,i+ii),
one can effectively obtain a rational sequence $(q_{\ell})$
whose set of accumulation points coincides with $A$.
\item[c)]
Given a rational sequence $(q_{\ell})$, a Turing machine can enumerate
a subset $Q$ of rational numbers having the same accumulation points.
(Recall that a sequence may repeat elements but a set cannot.)
\item[d)]
Given an enumeration of a subset $Q$ of rational numbers,
one can effectively generate a double sequence of open
rational intervals $\oball_{m,n}$ satisfying i+ii) above
where $A$ denotes the set of accumulation points of $Q$.
\item[e)]
If a double sequence of open rational intervals $\oball_{m,n}$ 
with i) is recursive, then the set $A$ according to ii)
is co-r.e. relative to $\emptyset'$.
\item[f)]
Let $N\in\IN$, $\vec u_n,\vec v_n\in\IQ^d$, and $\vec x\in\IR^d$ with 
$\vec x\not\in\bigcup_{n=1}^N (\vec u_n,\vec v_n)$.
Then, to every $\epsilon>0$,
there is some $\vec q\in\IQ^d\setminus\bigcup_{n=1}^N (\vec u_n,\vec v_n)$
such that $\|\vec x-\vec q\|\leq\epsilon$.%
\end{enumerate}
\end{lemma}
\begin{proof}
\begin{enumerate}
\item[a)]
 By \mycite{Lemma~5.1.10}{Weihrauch}, $\psig$--computability of a closed set
 $A$ implies (is even uniformly equivalent to)
 enumerability of open rational intervals $\oball_n$ with
 $A=\IR\setminus\bigcup_n \oball_n$. By application of 
 \textsf{Limit Lemma} 
 (\person{Shoenfield}'s? anyway, see for example \cite{Soare}) we conclude
 that $\psig$--computability \emph{relative} to $\emptyset'$
 implies (and follows from, see Item~e)
 recursivity of a double sequence $\oball_{m,n}$ satisfying i) and ii).
\item[b)]
 Calculate $\tilde N(m):=\max\{N\leq m:\oball_{m,n}=\oball_{m+1,n}\;\forall n\leq N\}$
 and let 
 $(q_{\langle m,k\rangle})_{_k}$
 enumerate (without repetition) the set 
 $\IQ\setminus\bigcup_{n=1}^{\tilde N(m)} \oball_{m,n}$.
\begin{itemize}
\item For $x\not\in A$,
 there exists $N$ such that $x\in \oball_{\infty,N}$.
 By i), $\tilde N(m)\geq N$ for $m\geq M(N)$.
 Therefore $q_{\langle m,k\rangle}\in \oball_{\infty,N}$ 
 can hold only for $m<M(N)$, i.e., finitely often;
 hence $x$ is no accumulation point of $(q_{\ell})$.
\item Suppose $x\in A$, i.e.
 $x\not\in\bigcup_n \oball_{\infty,n}\supseteq\bigcup_{n=1}^{N} \oball_{m,n}$
 for every $N$ and $m\geq M(N)$. In particular for $m\geq M(N)$,
 it holds $x\not\in\bigcup_{n=1}^{\tilde N(m)} \oball_{m,n}$ and 
 by construction plus Claim~f) 
 there is some $k_m$ with $|q_{\langle m,k_m\rangle}-x|\leq2^{-m}$.
 So $x$ is an accumulation point of $(q_{\ell})$.
\end{itemize}
\item[c)]
 Starting with $Q=\{\}$ add, inductively for each $\ell\in\IN$,
 a rational number not yet in $Q$ and closer to $q_\ell$ than $2^{-n}$.
 Indeed finiteness of $Q$ at each step
 asserts: $\exists p\in\IQ\cap\big((q_\ell-2^{-n},q_\ell+2^{-n})\setminus Q\big)$.
\item[d)]
 Let $(\oball_{0,k})_{_k}$ denote an effective enumeration of all open rational intervals.
 Given $(q_{\ell})_{_{\ell}}$, calculate inductively for $m\in\IN$ the subsequence
 $(\oball_{m+1,n})_{_n}$ of $(\oball_{m,n})_{_n}$ containing those intervals
 disjoint to $\{q_1,\ldots,q_m\}$.
\begin{itemize}
\item For $x$ accumulation point of $(q_\ell)$ and $\oball_{0,k}$ an arbitrary
 open rational interval containing $x$, there is some $q_M\in \oball_{0,k}$.
 By construction, this $\oball_{0,k}$ will not occur in $(\oball_{m+1,n})_{_n}$
 for $m\geq M$. Hence $x\in\IR\setminus\bigcup_n \oball_{\infty,n}=A$.
\item If $x$ is contained in some interval $\oball_{0,k}$
 which `prevails' as $\oball_{\infty,n}$,
 it cannot contain any $q_m$ by construction.
 Therefore $x$ is no accumulation point.
\end{itemize}
\item[e)]
 Consider a Turing machine enumerating $(\oball_{m,n})_{_{\langle m,n\rangle}}$.
 $(\oball_{M(N),N})_{_N}$ is a $\psig$--name of $A$
 \mycite{Lemma~5.1.10}{Weihrauch}.
 Deciding for given $N,M\in\IN$ whether ``$\oball_{m,n}=\oball_{M,n}$\;
 $\forall m\geq M$\; $\forall n\leq N$'' holds, is a $\Pi_1$--problem
 and thus possible relative to $\emptyset'$.
 With the help of this oracle, one can therefore
 compute $N\mapsto M(N)$ according to i).
\item[f)]
 If $\vec x\in\IQ^d$ then let $\vec q:=\vec x$.
 Otherwise $\vec x$ belongs to the open set $\IR\setminus\bigcup_{n=1}^N [\vec u_n,\vec v_n]$
 in which rational numbers lie dense.
\end{enumerate}\end{proof}

\section{Connected Components} \label{s:CC}
Instead of asking whether a set contains a computable point,
we now turn to the question whether it has a `computable' connected component.
Proofs here are more complicated but the general picture
turns out rather similar to Section~\ref{s:Closed}:
\begin{itemize}
\item[\textbullet] If the co-r.e. closed set under consideration contains
  finitely many components, each one is again co-r.e.
  (Section~\ref{s:CcFinite}).
\item[\textbullet] If there are countably many, some is co-r.e.
  (Section~\ref{s:CcCountable}).
\item[\textbullet] There exists a compact co-r.e. set of which
  none of its (uncountably many) connected components is co-r.e.
  (Observation~\ref{o:CC}).
\end{itemize}
Recall that for a topological space $X$,
the connected component $C(X,x)$ of $x\in X$ denotes
the union over all connected subsets of $X$ containing $x$.
It is connected and closed in $X$.
$C(X,x)$ and $C(X,y)$ either coincide or are disjoint.
\begin{proposition} \label{p:Connected} Fix $d\in\IN$.
\begin{enumerate}
\item[a)] Every (path\footnote{%
An open subset of Euclidean space is connected ~\myiff~ it is path-connected.}--)
connected component of an r.e. open set is r.e. open.
\\ More precisely (and more uniformly) the following mapping is well-defined and 
  $(\thetaL{d},\myrho^d,\thetaL{d})$--computable:
\[
\big\{(U,\vec x):\vec x\in U\subseteq\IR^d\text{ open}\big\}
\;\ni\; (U,\vec x) \;\mapsto\; C(U,\vec x)\; \subseteq\;\IR^d\text{ open.}
\]
\item[b)] 
The following mapping is well-defined and
  $(\psiG{d},\myrho^d,\psiG{d})$--computable:
\[
\big\{(A,\vec x):\vec x\in A\subseteq[0,1]^d\text{ closed}\big\}
\;\ni\; (A,\vec x) \;\mapsto\; C(A,\vec x)\;\subseteq\;[0,1]^d\text{ closed.}
\]
\end{enumerate}
\end{proposition}
\begin{proof}
First observe that closedness of $C(A,\vec x)$ in
closed $A\subseteq[0,1]^d$ means compactness in $\IR^d$. 
Similarly, open $U$ is locally (even path-) connected,
hence $C(U,\vec x)$ open in $U$ and thus also in $\IR^d$.
\begin{enumerate}
\item[a)] 
  Let $(B_1,B_2,\ldots,B_m,\ldots)$ denote a sequence of open 
  rational balls exhausting $U$, namely given as a $\thetaL{d}$--name of $U$.
  Since the non-disjoint union of two connected subsets is connected again,
\begin{equation} \label{e:Open} \vec x\in B_{m_1} \;\;\wedge\;\;
  B_{m_i}\cap B_{m_{i+1}}\not=\emptyset \;\; \forall i<n 
\end{equation}
  implies $B_{m_n}\subseteq C(U,\vec x)$ for any choice of
  $n,m_1,\ldots,m_n\in\IN$.
  Conversely, for instance by \mycite{Satz~4.14}{Querenburg}, there exists
  to every $\vec y\in C(U,\vec x)$ a finite subsequence 
  $B_{m_i}$ ($i=1,\ldots,n$) satisfying (\ref{e:Open})
  with $\vec y\in B_{m_n}$.
  Condition~(\ref{e:Open}) being semi-decidable, one can enumerate
  all such subsequences and use them to exhaust $C(U,\vec x)$.
  Nonuniformly, every connected component 
  contains by openness a rational (and thus computable) `handle' $\vec x$.
\item[b)]
  Recall the notion of a \emph{quasi-}component
  \mycite{\textsection46.V}{Kuratowski}
\begin{equation} \label{e:quasi1}
  Q(A,\vec x) \;:= \bigcap_{S\in\calS(A,\vec x)} \!\!\!S,
  \qquad \calS\;:=\;\big\{S\subseteq A\::\:S \text{ clopen in } A, \;\vec x\in S\big\}
\end{equation}
  where ``clopen in $A$'' means being both closed and open in the relative topology of $A$.
  That is, $S$ is closed in $\IR^d$, and so is $A\setminus S$!
  By the $T_4$ separation property (normal space), there exit
  disjoint open sets $U,V\subseteq\IR^d$ such that
  $S\subseteq U$ and $A\setminus S \subseteq V$.
  In particular $S=A\cap\closure{U}$, $U\cap V=\emptyset$, and $A\subseteq U\cup V$:
\begin{equation} \label{e:quasi2}
  \calS(A,\vec x) \;=\;\;\big\{ A\cap\closure{U}\;\big|\; U,V\subseteq\IR^d\text{ open},
   U\cap V=\emptyset,\: \vec x\in U,\: A\subseteq U\cup V\big\} \; .
\end{equation}
  Both $U$ and $V$ are unions from from the topological base of open rational balls;
  w.l.o.g. \emph{finite} such unions by compactness of $A$:
  $U=\closure{B_1\cup\ldots\cup B_n}=\cball_1\cup\ldots\cup\cball_n$ and
  $V=B_1'\cup\ldots\cup B_m'$. Therefore $Q(A,\vec x)$ coincides with
\begin{multline} \label{e:Closed}
 A\cap\bigcap\big\{\cball_1\cup\ldots\cup\cball_n\;\big|\;
 B_1,\ldots,B_n,B_1',\ldots,B_m'\text{ open rational balls}, \\
 B_i\cap B_j'=\emptyset,\: \vec x\in B_1,\: A\subseteq B_1\cup\ldots\cup B_m'\big\} \enspace .
\end{multline}
  Conditions ``$B_i\cap B_j'=\emptyset$'' and ``$\vec x\in B_1$'' are
  semi-decidable; and so is ``$A\subseteq B_1\cup\ldots\cup B_m'$'', 
  see for example \cite[\textsc{Lemma~4.1}b]{MLQ2}.
  Hence $Q(A,\vec x)$ is $\psiG{d}$--computable via
  the intersection (\ref{e:Closed}) by virtue of
  the countable variant of \mycite{Theorem~5.1.13.2}{Weihrauch},
  compare \mycite{Example~5.1.19.1}{Weihrauch}.
  Now finally, $Q(A,\vec x)=C(A,\vec x)$ since 
  components and quasi-components coincide for compact spaces
  \mycite{Theorem~\textsection47.II.2}{Kuratowski}.
\end{enumerate}\end{proof}
\noindent
Effective boundedness is essential in Proposition~\ref{p:Connected}b):
one can easily see that $A\mapsto C(A,\vec x)$ is in general 
$(\psiG{2},\psiG{2})$--discontinuous
for fixed computable $\vec x\in A$ 
when a bound on $A$ is unknown.
Non-uniformly, we have the following (counter-)
\begin{example} \label{x:CCH}
The following indicates an unbounded co-r.e. closed set $A\subseteq\IR^2$:\\
\centerline{\includegraphics[width=0.7\textwidth]{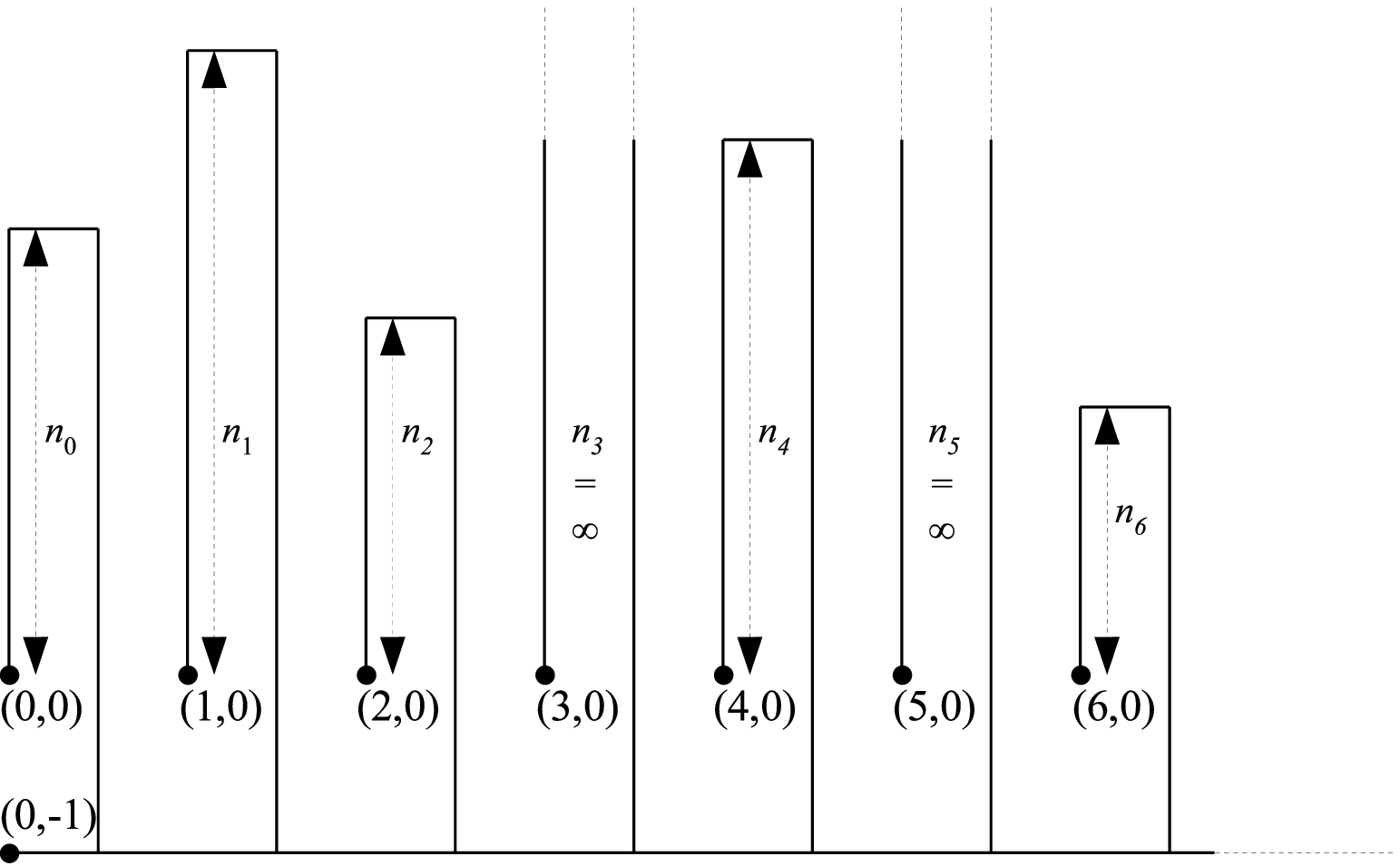}} \\
Here $n_e$ denotes the number of steps performed by the Turing machine 
with G\"{o}del index $e$ before termination (on empty input), 
$n_e=\infty$ if it does not terminate (i.e. $e\not\in H$). \\
Consider the connected component $C$ of $A$ with computable handle $(0,-1)$:
Were it co-r.e., then one could semi-decide ``$(e,0)\not\in C$''
\cite[\textsc{Lemma~4.1}c]{MLQ2},
equivalently: semi-decide ``$e\not\in H$'': contradiction.
\qed\end{example}
\noindent
As opposed to the open case a),
a computable `handle' $\vec x$ for a compact connected component $C(A,\vec x)$
need not exist; hence the non-uniform variant of b) may fail:%
\begin{observation} \label{o:CC}
A co-r.e. closed subset of $[0,1]$ obtained from Fact~\ref{f:Singular}
has uncountably many connected components, all singletons and none co-r.e.
\end{observation}
Indeed if $A\subseteq[0,1]$ has positive measure, it must contain
uncountably many points $x$. Each such $x$ is a connected component of
its own: otherwise $C(A,x)$ would be a non-empty interval and therefore
contain a rational (hence computable) element: contradiction.
%

Regarding that the counter-example according to Observation~\ref{o:CC} has
uncountably many connected components, it remains to study---in
analogy to Section~\ref{s:Nonuniform}---the cases of
countably infinitely many (Section~\ref{s:CcCountable})
and of
\subsection{Finitely Many Connected Components} \label{s:CcFinite}
Does every bounded co-r.e. closed set with \emph{finitely}
many connected components have a co-r.e. closed connected
component? Proposition~\ref{p:Connected}b) stays
inapplicable because there still need not exist a computable handle:

\begin{example}  \label{x:Stephane}
Let $A\subseteq[0,1]$ denote a non-empty co-r.e. closed set
without computable points (recall Fact~\ref{f:Singular}).
Then $(A\times[0,1])\cup([0,1]\times A)\subseteq[0,1]^2$
is (even \emph{path-}) \emph{connected} non-empty co-r.e. closed,
devoid of computable points.
\qed\end{example}
\noindent
Nevertheless, Proposition~\ref{p:Component}b+c) exhibits a
(partial) analog to Corollary~\ref{c:Finite}.
To this end, observe that a point $\vec x$ in some
set $A\subseteq\IR^d$ is isolated ~\myiff~ $\{\vec x\}$
is open in $A$.

\begin{proposition} \label{p:Component}
Let $\emptyset\not=A\subseteq\IR^d$ be closed.
\begin{enumerate}
\item[a)]
If $A$ has finitely many connected components,
then each such connected component is open in $A$.
\item[b)]
If $A$ is co-r.e. and $C(A,\vec x)$ a bounded connected
component of $A$ open in $A$, then $C(A,\vec x)$ is also co-r.e.
\item[c)]
If $A$ is r.e. and $C(A,\vec x)$ a bounded connected
component of $A$ open in $A$, then $C(A,\vec x)$ is also r.e.
\end{enumerate}
\end{proposition}

\begin{corollary} \label{c:Component}
If bounded co-r.e. closed $A\subseteq\IR^d$ has
only finitely many connected components,
then each of them is itself co-r.e.
\end{corollary}

\begin{proof}[Proof (Proposition~\ref{p:Component}).]
\begin{enumerate}
\item[a)]
Let $C(A,\vec x_1),\ldots,C(A,\vec x_k)$ denote the connected components of $A$.
Each of them is closed.
In particular, the finite union $C(A,\vec x_2)\cup\ldots\cup C(A,\vec x_k)$
is closed and equal to the complement of $C(A,\vec x_1)$ in $A$,
hence $C(A,\vec x_1)$ is also open in $A$.
\item[b)]
As $C(A,\vec x)$ is open in $A$,
there exists an open subset $U$ of $\IR^d$
such that $C(A,\vec x)=A\cap U$.
Closed $\IR^d\setminus U$ being disjoint from closed $C(A,\vec x)$, 
there are also disjoint open $V,W\subseteq\IR^d$
such that $C(A,\vec x)\subseteq V$ and 
$A\setminus C(A,\vec x)\subseteq\IR^d\setminus U\subseteq W$
($T_4$ separation property, normal space).
By compactness, there are finitely many open rational balls
$B_1,\ldots,B_n\subseteq V$ covering $C(A,\vec x)$.
Their centers and radii are in particular computable,
hence $\closure{B_1}\cup\ldots\cup\closure{B_n}$
is co-r.e. closed.
Moreover, as $\closure{B_1}\cup\ldots\cup\closure{B_n}\subseteq\closure{V}$ 
avoids $W\supseteq A\setminus C(A,\vec x)$,
it holds
$C(A,\vec x)=A\cap(\closure{B_1}\cup\ldots\cup\closure{B_n})$
which is co-r.e.
\item[c)]
As in b),
$C(A,\vec x)=A\cap(B_1\cup\ldots\cup B_n)$
is closed with r.e. open $B_1\cup\ldots\cup B_n$,
hence $C(A,\vec x)$ is itself r.e. 
by by Lemma~\ref{l:Intersect} below.
\end{enumerate}\end{proof}
\noindent
Although intersection of closed sets is in general discontinuous
\mycite{Theorem~5.1.13.3}{Weihrauch}, it holds%
\begin{lemma} \label{l:Intersect}
The following mapping is $(\psiL{d},\thetaL{d},\psiL{d})$--computable:
\[ \big\{(A,U):A\subseteq\IR^d\text{ closed},U\subseteq\IR^d\text{ open},
  A\cap U\text{ closed}\}
  \;\ni\; (A,U) \;\mapsto\; A\cap U \enspace . \]
\end{lemma}
\begin{proof}
Let $A$ be given as a sequence $(\vec x_n)_{_n}\subseteq A$ of real vectors
dense in $A$ \mycite{Lemma~5.1.10}{Weihrauch}.
Since ``$\vec x_n\in U$'' is semi-decidable \cite[\textsc{Lemma~4.1}c]{MLQ2},
one can effectively enumerate (possibly in different order)
all those $\vec x_n$ belonging to $U$.
Their closure thus lies in $\closure{A\cap U}$ which,
by presumption, coincides with $A\cap U$.
Conversely, to every $\vec z\in A\cap U$,
there exists some $\vec x_n\in U$ arbitrarily close to $\vec z$.
We thus conclude that the output subsequence of $(\vec x_n)$ is 
(equivalent to) a valid $\psiL{d}$--name of $A\cap U$.
\end{proof}

\subsection{Countably Infinitely Many Connected Components} \label{s:CcCountable}
By Proposition~\ref{p:Component}a+b), if bounded co-r.e. closed $A\subseteq\IR^d$ 
has finitely many components, each one is itself co-r.e. 
In the case of countably infinitely many connected components,
we have seen in Example~\ref{x:CCH} a bounded co-r.e. closed set 
containing a connected component which is not co-r.e.;
others of its components on the other hand are co-r.e.
In fact it holds the following counterpart to Fact~\ref{f:Isolated}b):

\begin{lemma} \label{l:perfect}
Let $\emptyset\not=A\subseteq\IR^d$ be compact with no
connected component open in $A$. Then $A$ has as many connected
components as cardinality of the continuum.
\end{lemma}
\noindent
Proposition~\ref{p:Component}b) implies
\begin{corollary} \label{c:perfect}
Let $A\subseteq\IR^d$ be compact and co-r.e. with countable many
connected components. Then at least one such component is again co-r.e.
\end{corollary}

\begin{proof}[Proof (Lemma~\ref{l:perfect}).]
By \mycite{Theorem~\textsection46.V.3}{Kuratowski},
there exists a continuous function $f:A\to\{0,1\}^\omega$
such that the point inverses $f^{-1}(\bar\sigma)$ coincide
with the quasi-components of $A$; and these in turn with $A$'s 
connected components \mycite{Theorem~\textsection47.II.2}{Kuratowski}.
Since $A$ is compact and $f$ continuous, $f[A]\subseteq\{0,1\}^\omega$ is compact, too.
Moreover every isolated point $\{\bar\sigma\}$ of $f[A]$ yields
$f^{-1}(\bar\sigma)$ (closed and) open a component in $A$.
So if $A$ has no open component, $f[A]$ must be perfect---and
thus of continuum cardinality by virtue of Fact~\ref{f:Isolated}b).
\end{proof}
Corollary~\ref{c:perfect} and Example~\ref{x:CCH} leave open the following
\begin{question}
Is there a \emph{bounded} co-r.e. closed set with countably 
many connected components, one of which is \emph{not} co-r.e.?
\end{question}
In view of Proposition~\ref{p:Connected}b),
this component must not contain a computable point.

\subsection{Related Work}
An anonymous referee has directed our attention to the following 
interesting result which appeared as \mycite{Theorem~2.6.1}{Miller}:
\begin{fact} \label{f:Miller}
For any co-r.e. closed $X\subseteq[0,1]^d$, the following are equivalent:
\begin{enumerate}
\item[(1)] $X$ contains a nonempty co-r.e. closed connected component,
\item[(2)] $X$ is the set of fixed points of some computable map $g:[0,1]^d\to[0,1]^d$,
\item[(3)] the image $f(X)$ contains a computable number for any computable $f:X\to\IR$.
\end{enumerate}
\end{fact}

\section{Co-R.E. Closed Sets \emph{with} Computable Points}
The co-r.e. closed subsets of $\IR$ devoid of computable points according to
Fact~\ref{f:Singular} lack convexity:

\begin{observation} \label{o:interval}
Every non-empty co-r.e. interval $I\subseteq\IR$ trivially
has a computable element:

Either $I$ contains an open set (and thus lots of rational elements $x\in I$)
or it is a singleton $I=\{x\}$, hence $x$ computable \mycite{Proposition~3.6}{Closed}.
\end{observation}
(It is not possible to continuously `choose', even in a multi-valued way, 
some $x\in I$ from a $\psig$--name of $I$, though\ldots)
This generalizes to higher dimensions:

\begin{theorem} \label{t:Convex}
Let $\emptyset\not=A\subseteq\IR^d$ be co-r.e. closed and convex.
Then there exists a computable point $\vec x\in A$.
\end{theorem}
\begin{proof}
W.l.o.g. suppose $A$ is compact by intersection
with some sufficiently large cube $[-N,+N]^d$.
Then proceed by induction on $d$:
Under projection $(x_1,\ldots,x_{d-1},x_d)\mapsto x_d$,
the image 
\[ A_d \quad:=\quad \big\{ x_d \;\big|\; \exists x_1,\ldots,x_{d-1}\in\IR:
  (x_1,\ldots,x_{d-1},x_d)\in A\big\} \quad\subseteq\;\IR \]
is convex and
$\psig$--computable by virtue of \mycite{Theorem~6.2.4.4}{Weihrauch},
hence contains a computable point $x_d\in A_d$ (Observation~\ref{o:interval}).
The intersection $A\cap (\IR^{d-1}\times\{x_d\})$
is therefore non-empty, also convex, and $\psiG{d-1}$--computable
\mycite{Theorem~5.1.13.2}{Weihrauch};
hence it contains a computable point $(x_1,\ldots,x_{d-1})$
by induction hypothesis.
Then $(x_1,\ldots,x_{d-1},x_d)$ is a computable element of $A$.
\end{proof}

\subsection{Star-Shaped Sets}
A common weakening of convexity is given in the following
\begin{definition}\it \label{d:star}
A set $A\subseteq\IR^d$ is \emph{star-shaped} if there exists
a (so-called \emph{star-}) point $\vec s\in A$ such that,
for every $\vec a\in A$, the line segment\footnote{The reader is not
in danger of confusing this with the same notion 
$[\vec s,\vec a]$ standing for
the cube $\prod_i [s_i,a_i]$ in Sections~\ref{s:Closed} and \ref{s:Naive}.}
$[\vec s,\vec a]:=\{\lambda\vec s+(1-\lambda)\vec a:0\leq \lambda\leq1\}$
is contained in $A$.

The \emph{set of star-points} $S(A)$ is the collection
of \emph{all} star-points of $A$.
\end{definition}
\noindent
So $A$ is convex ~\myiff~ $A=S(A)$;
$A$ is star-shaped ~\myiff~ $S(A)\not=\emptyset$;
and star-shape implies (even simply-)connectedness.
\begin{figure}[htb]
\includegraphics[width=\textwidth]{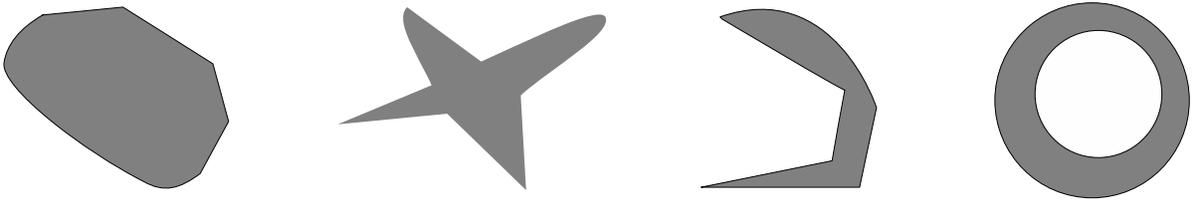}%
\caption{\label{f:convex}A convex, a star-shaped,
a simply-connected, and a connected set.}
\end{figure}
\begin{lemma} \label{l:star}
$S(A)\subseteq A$ is convex. Moreover if $A$ is closed, then so is $S(A)$.
\end{lemma}
\begin{proof}
Let $\vec x,\vec y$ be star-points of $A$ and $\vec a\in A$ arbitrary.
By prerequisite the three segments $[\vec x,\vec a]$ and $[\vec y,\vec a]$
and $[\vec x,\vec y]$ all lie in $A$. 
Moreover each segment $[\vec x,\vec b]$ with $\vec b\in[\vec y,\vec a]$---that 
is the entire closed triangle spanned by $(\vec x,\vec y,\vec b)$---also 
belongs to $A$; in particular each segment $[\vec c,\vec a]$ with
$\vec c\in[\vec x,\vec y]$ does. Since $\vec a\in A$ was arbitrary,
this asserts each such $\vec c$ to be a star-point of $A$.

Let $(\vec x_n)$ be a sequence of star-points converging to some $\vec x\in A$.
For arbitrary $\vec a\in A$, the segments $[\vec x_n,\vec a]$ all belong to 
closed $A$, hence so does $[\vec x,\vec a]$.
\end{proof}

\begin{theorem} \label{t:LeRoux}
Let $\emptyset\not=A\subseteq\IR^2$ be co-r.e. closed and star-shaped.
Then $A$ contains a computable point.
\end{theorem}
\noindent
In view of Lemma~\ref{l:star} this claim would follow from Theorem~\ref{t:Convex}
if, for every star-shaped co-r.e. closed $A$, 
its set $S(A)$ of star-points were co-r.e. again.
However we have been shown the latter assertion to fail already 
for very simple compact subsets in 2D \cite{Miller2}.

\begin{figure}[htb]
\includegraphics[width=\textwidth]{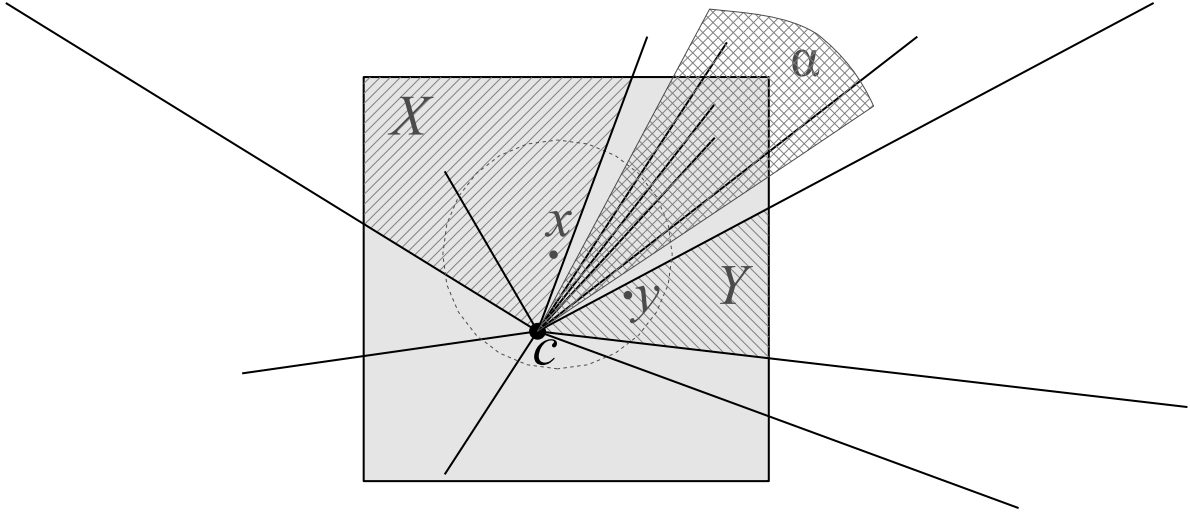}
\caption{\label{f:star}Illustration to the proof of Theorem~\ref{t:LeRoux}
for the case $S(A)=\{\vec c\}$.}
\end{figure}

\begin{proof}[Proof (Theorem~\ref{t:LeRoux}).]
If $A$ has non-empty interior, it contains a rational (and thus computable) point.
Otherwise suppose the convex set $S(A)$ to have dimension one,
i.e. $S(A)=[\vec x,\vec y]$ with distinct $\vec x,\vec y\in A$.
Were $S(A)$ a \emph{strict} subset of $A$, $A$ would contain
an entire triangle (compare the proof of Lemma~\ref{l:star}) 
contradicting $A^\circ=\emptyset$. Hence $S(A)=A$ is co-r.e. 
and contains a computable point by Theorem~\ref{t:Convex}.

It remains to treat the case of $S(A)=\{\vec c\}\subsetneq A$,
$A$ consisting of semi-/rays originating from $\vec c$
as indicated in Figure~\ref{f:star}.
Consider some rational square $Q$ containing $\vec c$ in its interior
but not the entire $A$. If the square's boundary, intersected with $A$,
contains an isolated point, this point will be computable
according to \mycite{Theorem~5.1.13.2}{Weihrauch} and
Section~\ref{s:Nonuniform}. Otherwise $Q^\circ\setminus A$ consists
of uncountably many (Observation~\ref{o:Main}) connected components.
Let $X$ and $Y$ denote two \emph{non-}adjacent ones of them,
each r.e. open according to Proposition~\ref{p:Connected}a).
Also let $0<\alpha\leq180^\circ$ be some (w.l.o.g. rational and thus computable) 
lower bound on the angle between $X$ and $Y$. Notice that 
$X$ and $Y$ `almost touch' (i.e. their respective closures meet)
exactly in the sought point $\vec c$.
Moreover for $\vec x\in X$ and $\vec y\in Y$, elementary trigonometry confirms that
$\|\vec x-\vec c\|_2\leq \|\vec x-\vec y\|_2/(2\sin\tfrac{\alpha}{2})$.
Based on effective enumerations of all rational $\vec x\in X$ and all rational $\vec y\in Y$,
we thus obtain arbitrary good approximations to $\vec c$.
\end{proof}

Regarding a further weakening from convex over star shape, we ask
\begin{question} \label{q:Miller}
Does every simply-connected co-r.e. closed non-empty subset
of $[0,1]^2$ contain a computable point?
\end{question}
Mere connectedness is not sufficient: recall Example~\ref{x:Stephane}.
This immediately extends to a (counter-)example 
giving a negative answer to Question~\ref{q:Miller} in 3D:
\begin{example}
Let $A\subseteq[0,1]$ denote a non-empty co-r.e. closed set
without computable points.
Then $(A\times[0,1]^2)\cup([0,1]\times A\times[0,1])\cup([0,1]^2\times A)\subseteq[0,1]^3$
is simply-connected non-empty co-r.e. closed
devoid of computable points.
\qed\end{example}



\begin{thebibliography}{[WeZh00}
\bibitem[Bees85]{Beeson}
  \textsc{M.J.~Beeson}: ``\emph{Foundations of Constructive Mathematics}'',
  Springer (1985).
\bibitem[Boto79]{Querenburg}
  \textsc{B.von Querenburg}: ``\emph{Mengentheoretische Topologie}'',
  Springer (1979).
\bibitem[Brat05]{Borel}
  \textsc{V.~Brattka}: ``Effective Borel Measurability and Reducibility of Functions'',
  pp.19--44 in \emph{Mathematical Logic Quarterly} vol.\textbf{51:1} (2005).
\bibitem[BrWe99]{Closed}
  \textsc{V.~Brattka, K.~Weihrauch}: ``Computability on
  Subsets of Euclidean Space I: Closed and Compact Subsets'', pp.65-93
  in \emph{Theoretical Computer Science} vol.\textbf{219} (1999).
\bibitem[CeRe98]{Cenzer} 
  \textsc{D.~Cenzer, J.B.~Remmel}: ``$\Pi^0_1$ Classes in Mathematics'',
  pp.623--821 in \textsc{Yu.L.~Ershov, S.S.~Goncharov, A.~Nerode, J.B.~Remmel} (Eds.)
  \emph{Handbook of Recursive Mathematics} vol.\textbf{2}, Elsevier (1998).
\bibitem[GeNe94]{Ge} 
  \textsc{X.~Ge, A.~Nerode}: ``On Extreme Points of Convex
  Compact Turing Located Sets'', pp.114-128 in \emph{Logical
  Foundations of Computer Science}, Springer LNCS vol.\textbf{813} (1994).
\bibitem[Gher06]{Gherardi}
  \textsc{G.~Gherardi}: ``An Analysis of the Lemmas of Urysohn
  and Urysohn-Tietze According to Effective Borel Measurability'',
  pp.199--208 in \emph{Proc. 2nd Conference on Computability in Europe} 
  (CiE'06), Springer LNCS vol.\textbf{3988}.
\bibitem[Ho99]{Ho}
  \textsc{C.-K.~Ho}: ``Relatively recursive reals and real functions'',
  pp.99--120 in \emph{Theoretical Computer Science} vol.\textbf{210} (1999).
\bibitem[Kech95]{Kechris}
  \textsc{A.S.~Kechris}: ``\emph{Classical Descriptive Set Theory}'',
  Springer (1995).
\bibitem[KrLa57]{Kreisel}
  \textsc{G.~Kreisel, D.~Lacombe}: ``Ensembles r\'{e}cursivement measurables
  et ensembles r\'{e}cursivement ouverts ou ferm\'{e}s'', pp.1106--1109 in
  \emph{Compt.\ Rend.\ Acad.\ des Sci.\ Paris} vol.\textbf{245} (1957).
\bibitem[Kura68]{Kuratowski}
  \textsc{K.~Kuratowski}:  ``\emph{Topology Vol.II}'',
  Academic Press (1968).
\bibitem[Kush84]{Kushner}
  \textsc{B.~Kushner}: ``\emph{Lectures on Constructive Mathematical Analysis}'',
  vol.\textbf{60}, American Mathematical Society (1984).
\bibitem[Laco57]{LacombeI}
  \textsc{D.~Lacombe}: ``Les ensembles r\'{e}cursivement ouverts ou ferm\'{e}s,
  et leurs applications \`{a} l'analyse r\'{e}cursive I'',
  pp.1040--1043 in \emph{Compt.\ Rend.\ Acad.\ des Sci.\ Paris} vol.\textbf{245} (1957).
\bibitem[Laco58]{LacombeII}
  \textsc{D.~Lacombe}: ``Les ensembles r\'{e}cursivement ouverts ou ferm\'{e}s,
  et leurs applications \`{a} l'analyse r\'{e}cursive II'',
  pp.28--31 in \emph{Compt.\ Rend.\ Acad.\ des Sci.\ Paris}, vol.\textbf{246} (1958).
\bibitem[Lagn06]{Lagnese}
  \textsc{G.~Lagnese}: ``Can someone give me an example of\ldots'',
  in \texttt{Usenet}
  {\sf http://cs.nyu.edu/pipermail/fom/2006-February/009835.html}
\bibitem[Mill02]{Miller}
  \textsc{J.S.~Miller}: ``Pi-0-1 Classes in Computable Analysis and Topology'',
  PhD thesis, Cornell University, Ithaca, USA (2002).
\bibitem[Mill07]{Miller2}
  \textsc{J.S.~Miller}, Personal Communication (June 21, 2007).
\bibitem[Morr69]{Dictionary}
  \textsc{W.~Morris} (Editor): ``\emph{American Heritage Dictionary
  of the English Language}'', American Heritage Publishing (1969).
\bibitem[Spec59]{Specker}
  \textsc{E.~Specker}: ``Der Satz vom Maximum in der rekursiven Analysis'',
  pp.254--265 in \emph{Constructivity in Mathematics} (A.~Heyting Edt.),
  Studies in Logic and The Foundations of Mathematics, North-Holland (1959).
\bibitem[Soar87]{Soare}
  \textsc{R.I.~Soare}: ``\emph{Recursively Enumerable Sets and Degrees}'',
\bibitem[Weih00]{Weihrauch}
  \textsc{K.~Weihrauch}: ``\emph{Computable Analysis}'', Springer (2000).
\bibitem[WeZh00]{SemiTCS}
  \textsc{K.~Weihrauch, X.~Zheng}:
  ``Computability on continuous, lower semi-continuous and
  upper semi-continuous real functions'', pp.109--133 in
  \emph{Theoretical Computer Science} vol.{234} (2000).
\bibitem[ZaTs62]{Zaslavski}
  \textsc{I.D.~Zaslavski\u{\i}, G.S.~Tse\u{\i}tin}:
  ``On singular coverings and related properties of constructive functions'',
  pp.458--502 in \emph{Trudy Mat. Inst. Steklov.} vol.\textbf{67} (1962);\\
  English transl. in \emph{Amer. Math. Soc. Transl.} (2) \textbf{98} (1971).
\bibitem[Zieg04]{MLQ2}
  \textsc{M.~Ziegler}: ``Computable operators on regular sets'',
  pp.392--404 in \emph{Mathematical Logic Quarterly} vol.\textbf{50} (2004).
\bibitem[Zhen07]{Xizhong} 
  \textsc{X.~Zheng}, Personal Communication (June 21, 2007).
\bibitem[ZhWe01]{ArithHierarchy}
  \textsc{X.~Zheng, K.~Weihrauch}: ``The Arithmetical Hierarchy of Real Numbers'',
  pp.51--65 in \emph{Mathematical Logic Quarterly} vol.\textbf{47:1} (2001).
\end{thebibliography}
\end{document}